\documentstyle[twocolumn,aps]{revtex}
\newcommand{\be}{\begin{equation}} 
\newcommand{\ee}{\end{equation}} 
\newcommand{\bea}{\begin{eqnarray}} 
\newcommand{\eea}{\end{eqnarray}}

\begin{document} 
\draft 
\title{Boundary States and Black Hole Entropy}  
\author{J. David Brown} 
\address{Department of Physics and Department of Mathematics,\\ 
   North Carolina State University, Raleigh, NC 27695--8202}
\maketitle 
\begin{abstract} 
Black hole entropy is derived from a sum over boundary states.  The boundary 
states are labeled by energy and momentum surface densities, and parametrized 
by the boundary metric. The sum over state labels is expressed as a 
functional integral with measure  determined by the density of states. 
The sum over metrics is expressed as a functional integral with measure 
determined  by the universal expression for  the inverse temperature 
gradient at the horizon. The analysis applies to  any stationary, nonextreme 
black hole in any theory of gravitational and  matter fields. 
\end{abstract} 
Many researchers have suggested that black hole entropy arises 
from a sum over boundary (or horizon or surface or edge) states 
\cite{Bstates}. The formal analysis presented in this paper shows 
that this conjecture is essentially correct. 
I consider a system of gravitational and matter fields with 
outer and inner boundary elements, denoted $B$ and $B'$ respectively. 
The entropy for stationary, nonextreme black holes is derived by summing 
over all states  on the inner boundary element.  The states on $B'$ are 
labeled by the energy surface density $\varepsilon'$
and momentum surface  density $j_i'$ \cite{QLEandMFI}, and parametrized by 
the surface metric $\sigma'_{ab}$. The  measure for the integration  over 
$\varepsilon'$ and $j_i'$ is determined  by  the density of
states,  which is calculated formally as a path integral.  Integration 
over $\varepsilon'$ and $j_i'$ imposes the condition that $B'$ is  
the bifurcation surface of a 
stationary  black hole horizon. The measure for the integration over 
$\sigma_{ab}'$ is not fixed by the density of states  and requires 
further physical input for its determination. The needed  physical 
input is the inverse temperature gradient at
the horizon.  The inverse temperature gradient is a universal quantity, 
in the sense 
that it is independent of the details of the black hole and independent 
of the theory of gravity under consideration. In this regard, its input 
into the calculation is analogous to the identification of inverse 
temperature with imaginary time period. 

Recent results indicate that black hole entropy can be understood in terms 
of D--brane states in string theory \cite{Dbranes}. Although the string 
theory description of black hole entropy is quite compelling, it is 
limited in scope to a particular theory of gravity (as defined by the 
low energy limit of string theory). On the other hand, path integral and 
variational methods have shown that black hole entropy appears in all 
diffeomorphism invariant theories of gravitational and matter fields 
\cite{IandW,EntropyInGeneral}. This  suggests that a deeper understanding 
of black hole entropy can be found  from within the path integral formalism 
itself, without specialization  to a particular theory of gravity. The 
analysis contained in this paper  points to such a deeper understanding 
by providing an interpretation of  the path integral derivation of black 
hole entropy as a sum over boundary  states. 

Let me begin by presenting a quick overview of certain key aspects of 
statistical mechanics. Consider a nongravitating, quantum mechanical 
system with
Hamiltonian  operator ${\hat H}_V$ which depends on, say, volume $V$. 
(For example, the electromagnetic field in a 
spherical container of volume $V$.) The  partition function is 
defined by $Z_V(\beta) = {\rm tr}\, e^{-\beta {\hat H}_V}$, where $\beta$ 
is the inverse temperature.  Eigenstates of ${\hat H}_V$ are labeled by the 
corresponding eigenvalue, namely, energy $E$, and are parametrized by  
volume $V$. The partition function can be written as  
\begin{equation} 
  Z_V(\beta) = \int dE\,D_V(E)\, e^{-\beta E} \ ,\eqnum{1}
\end{equation}
where $D_V(E)$ is the density of energy eigenstates. That is, 
$D_V(E) \,dE$ is the number of states with volume $V$ and energy in the 
range $dE$. An alternative proceedure is to define $Z_V(\beta)$ by a 
path integral over histories that are periodic in imaginary time with 
period $\beta$. The phase in the path integral is given by the  
action, which is related to the classical Hamiltonian $H_V$ in 
the usual way.
The density of states $D_V(E)$ becomes a path integral over 
histories with fixed energy $E$, where $E$ is the value of the classical 
Hamiltonian \cite{QLEandMFI}. 

For a given value of $\beta$, the probability that the system will be 
found in a state with volume $V$ and energy in the  range $dE$ is 
${\cal P} D_V(E)\, dE$, where ${\cal P} = e^{-\beta E}/Z_V(\beta)$. The 
expectation value of energy is defined by 
$\langle E\rangle = - \partial(\ln Z_V(\beta))/\partial\beta$, pressure 
is defined by $\beta p = \partial(\ln Z_V(\beta))/\partial V$, and 
entropy is defined by 
\begin{equation} 
   {\cal S}_V(\beta)  =  - \int dE\, D_V(E)\, {\cal P}\ln{\cal P} 
     =  \ln Z_V(\beta) + \beta \langle E\rangle \ .\eqnum{2}
\end{equation} 
This is the entropy of the system in the canonical ensemble at temperature 
$1/\beta$. The microcanonical entropy is obtained from the infinite 
temperature limit, $\beta=0$, in which ${\cal P}$ is constant. In this 
limit Eqs.~(1) and (2) yield ${\cal S}_V(0) = 
\ln ( \int dE\, D_V(E) )$, which is the logarithm of the number of states 
with volume $V$. The total number of states $N$, with no restriction on  
$V$, is given by 
\begin{equation} 
   N = \int dE\,dV\,\mu(V) \, D_V(E)  \ .\eqnum{3}
\end{equation} 
Here, $\mu(V)$ is a measure for the $V$ integration that must be specified 
by some physical criterion. This unspecified measure 
also appears in the definition of the  partition function for the 
constant pressure ensemble \cite{SachHill}, namely, 
$Z_p(\beta) = \int dV\mu(V) e^{-\beta p V} Z_V(\beta)$. In fact, one can 
view $N$ as the infinite temperature limit of $Z_p(\beta)$. 

The density of states factor in Eq.~(1) can be written as $e^{\ln(D_V(E))}$, 
where for simplicity an arbitrary dimensionful constant (such as the Planck 
energy) has been omitted from the logarithm. Let us  evaluate the integral 
for $Z_V(\beta)$ in the steepest descents approximation. With $E=E^*(\beta)$ 
denoting the solution of
the  extremum condition $\beta = \partial( \ln D_V(E)) /\partial E$, we find 
\begin{equation} 
    \ln Z_V(\beta) \approx -\beta E^* + \ln [D_V(E^*) \, \Delta] 
    \eqnum{4} 
\end{equation} 
where $2\pi \Delta^{-2} = - \left.[\partial^2 (\ln D_V(E)) 
/\partial (E)^2 ]\right|_{E^*}$. The steepest descents approximation 
also yields $\langle E\rangle \approx E^*$. By differentiating the relation 
$\beta \approx \left.[\partial( \ln Z_V(E)) 
/\partial E]\right|_{\langle E\rangle}$ with respect to $\beta$, we find 
that $\Delta^2$ is given by the mean square deviation in energy, 
$\Delta ^2/(2\pi) \approx \langle ( E - \langle E \rangle)^2 \rangle$. 
A comparison of Eqs.~(2) and (4) shows that 
\begin{equation} 
    {\cal S}_V(\beta) \approx \ln [ D_V(E^*) \,\Delta] \ .\eqnum{5} 
\end{equation} 
Thus, the entropy in the canonical ensemble at temperature 
$1/\beta$ is the logarithm of the number of states in an energy interval 
given by the mean square deviation in energy. Equation (5) 
can be written as ${\cal S}_V(\beta) \approx \ln[ Z_V(E)\,\Delta ]$ where 
$\beta$ and $E$ are related by 
$\beta= \partial( \ln D_V(E)) /\partial E$. 

Now consider vacuum Einstein gravity 
on a spatial manifold $\Sigma$. The boundary $\partial\Sigma$ need not be 
simply connected---later we will specialize to the case in which 
$\partial\Sigma$ consists of an inner boundary element $B'$ and an outer 
boundary element $B$. The partition function is defined as a path 
integral on the manifold $\Sigma \times S^1$ with action\footnote{See 
Ref.~\cite{QLEandMFI}. The action in Eq.~(6) is 
defined by the exponential in the path integral, and therefore equals 
$i$ times the ``Lorentzian action". Also, for simplicity, the condition 
$V^i n_i = 0$ on $\partial\Sigma$ has been imposed;  
this means that the boundary is at rest with respect to the constant 
time slices. The action in the presence of moving boundaries is given 
in Ref.~\cite{BoostedQLE}.}
\begin{eqnarray} 
   S^{pf} & = & i\int dt \left( \int_\Sigma d^3x\, P^{ij} {\dot h}_{ij} 
   - H_\sigma \right)  \ ,\eqnum{6a}\\
   H_\sigma & = & \int_{\Sigma} d^3x ( N{\cal H} + V^i{\cal H}_i) \nonumber\\
    & + & \int_{\partial\Sigma} d^2x \sqrt{\sigma} (N\varepsilon - V^i j_i) 
   \ ,\eqnum{6b}
\end{eqnarray} 
where $h_{ij}$ and $P^{ij}$ are the canonical variables, $N$ and $V^i$ are the
lapse function and shift vector, and ${\cal H}$ and 
${\cal H}_i$ are the Hamiltonian and momentum constraints. Also, $\sigma_{ab}$ 
is the metric on $\partial\Sigma$ and 
$\varepsilon = k/(8\pi)$ and $j_i = -2P_{ij}n^j/\sqrt{h}$ are the energy surface 
density and momentum surface density \cite{QLEandMFI}. Here, 
$k= \sigma^{ab}k_{ab}$ and $n^j$ denote, respectively, the trace of the 
extrinsic curvature and the outward pointing unit normal of $\partial\Sigma$ 
embedded in $\Sigma$.  

The  variation of the action (6) is \cite{QLEandMFI} 
\begin{eqnarray} 
   \delta S^{pf} = \cdots 
   - i\int dt\int_{\partial\Sigma}d^2x \,\sqrt{\sigma} & &
   \{\varepsilon \delta N - j_i \delta V^i \nonumber\\ 
   & & - (Ns^{ab}/2)\delta\sigma_{ab}\} 
   \ ,\eqnum{7}
\end{eqnarray}
where the unwritten terms yield the classical equations of motion 
and $(Ns^{ab}/2) = [Nk^{ab} + (n^i\partial_i N - Nk)\sigma^{ab}]/(16\pi)$ 
defines the spatial stress. 
The form of $\delta S^{pf}$ shows that $\sigma_{ab}$ and the boundary values 
of $N$ and $V^i$ are held fixed in the variational principle. The partition 
function, defined by the  path 
integral, is a functional of these quantities. The boundary value of 
$N$ determines the period in imaginary time, which is identified with the 
inverse temperature on the system boundary: $\beta = 
i\int \left. dt\,N \right|_{\partial\Sigma}$. (Note that $\beta$ is not 
typically constant on $\partial\Sigma$ due to gravitational redshifting 
and blueshifting.)  The 
boundary  value of $V^i$ defines the proper velocity $v^i$ of the spatial 
coordinate system through the relation 
$\beta v^i = i\int \left. dt\, V^i\right|_{\partial\Sigma}$. 
(In the classical approximation, $v^i$ is the spatial velocity of 
the system with respect to  observers at rest on the boundary 
\cite{BMY,QLEandMFI}.) The partition function will be denoted 
$Z_\sigma[\beta,v]$. To be precise, 
$Z_\sigma[\beta,v]$ is the grand canonical partition function 
and $v^i$ plays the role of the ``chemical potential" for $j_i$. 

Because the Hamiltonian and momentum constraints vanish on shell, the value 
of the Hamiltonian $H_\sigma$ for various choices of lapse and shift is 
given by the densitized energy surface density $\sqrt{\sigma}\varepsilon$ 
and the densitized momentum surface density $\sqrt{\sigma}j_i$. These 
quantities play the role of energy. The fact that there are many such 
``energies" reflects the many--fingered character of time in gravitational 
physics. The fact  that they are
surface quantities reflects the fact that  the gravitational field on 
$\partial\Sigma$ is determined by the mass--energy throughout $\Sigma$. 
By analogy with the nongravitating system discussed above, we are lead 
to the formal identification of $\sqrt{\sigma}\varepsilon$ and 
$\sqrt{\sigma}j_i$ as labels for (boundary) states. These states are 
parametrized by the boundary metric $\sigma_{ab}$. 
In direct analogy with Eq.~(1), the partition function can be 
written as 
\begin{equation} 
   Z_\sigma[\beta,v] = \int d\varepsilon\,dj\,D_\sigma[\varepsilon,j] 
   e^{-\int_{\partial\Sigma}d^2x\,\sqrt{\sigma} 
   (\beta\varepsilon - \beta v^i j_i)} 
   \ , \eqnum{8} 
\end{equation} 
where  $\int d\varepsilon \,dj$ denotes functional integration over 
$\sqrt{\sigma}\varepsilon$ and $\sqrt{\sigma}j_i$. In Eq.~(8), 
$D_\sigma[\varepsilon,j]$ is the density of boundary states; that 
is, $D_\sigma[\varepsilon,j] d\varepsilon \,dj$ is the number of boundary 
states with metric $\sigma_{ab}$ and energy and momentum surface 
densities in the range $d\varepsilon \,dj$. 

The entropy of the gravitational field in the canonical ensemble with 
boundary temperature $1/\beta$ can be computed from the partition 
function $Z_\sigma[\beta,v]$ using the analog of Eq.~(2). Alternatively, 
one can use the density of states $D_\sigma[\varepsilon,j]$ and the 
analog of Eq.~(5). In the zero--loop approximation, in which 
only the dominant exponential contributions to the path integrals 
for $Z_\sigma[\beta,v]$ and $D_\sigma[\varepsilon,j]$ are kept, 
the second option is easier. Indeed, in this approximation the mean 
square deviation factor $\Delta$ can be dropped and the analog of Eq.~(5) 
becomes ${\cal S}_\sigma[\beta,v] \approx \ln(D_\sigma[\varepsilon,j])$. 
Here, it is understood that 
$\beta = \delta(\ln D_\sigma[\varepsilon,j])/\delta(\sqrt{\sigma}\varepsilon)$ 
with a similar relation for $\beta v^i$.  Now, from the path integral  
for $Z_\sigma[\varepsilon,j]$, we find that the density of states is 
expressed as a path
integral over the manifold  $\Sigma\times S^1$  with action   
\begin{equation} 
   S^{ds} = i \int dt\int_{\Sigma} d^3x ( P^{ij} {\dot h}_{ij} - N{\cal H} - 
   V^i{\cal H}_i ) \ .\eqnum{9} 
\end{equation} 
In the zero--loop approximation, $D_\sigma[\varepsilon,j]$ equals the 
exponential of the action $S^{ds}$ evaluated at the stationary classical 
solution (if one exists) with the given values of $\varepsilon$, $j_i$, 
and $\sigma_{ab}$ on $\partial\Sigma$. Since $S^{ds} = 0$ for a 
stationary classical solution, the zero--loop contribution to 
$D_\sigma[\varepsilon,j]$ is $1$ and the entropy 
${\cal S}_\sigma[\beta,v] \approx \ln(D_\sigma[\varepsilon,j])$ vanishes. 

The conclusion ${\cal S}_\sigma[\beta,v] \approx 0$ is general. 
Note, however, that $\varepsilon$ and $j_i$ label the states 
for the complete boundary 
$\partial\Sigma$ of the spatial manifold. If $\partial\Sigma$ 
consists of disconnected pieces, then  $\varepsilon$ and $j_i$ include 
labels for states on each boundary element. Likewise, 
$\beta$, $v^i$, and $\sigma_{ab}$ specify  the inverse temperature, 
velocity, and metric on each boundary element. 
Let us  assume that $\partial\Sigma$ consists of an 
``outer" boundary  element $B$ and an ``inner" boundary element $B'$, and 
make this dependence  explicit by writing 
$D[\varepsilon,j;\sigma| \varepsilon',j';\sigma']$  for the density of 
states and $Z[\beta,v;\sigma| \beta',v';\sigma']$ for the  partition function. 
Here, the unprimed quantities refer to $B$ and the primed 
quantities refer to $B'$. The vanishing entropy is expressed as 
${\cal S}[\beta,v;\sigma| \beta',v';\sigma'] \approx 0$. 

As an example, consider a Schwarzschild black hole. Let the data 
$\varepsilon$, $j_i$, $\sigma_{ab}$ on $B$ and the data $\varepsilon'$, 
$j_i'$, $\sigma_{ab}'$ on $B'$ coincide with those for two 
``$r={\rm const}$" surfaces. By construction, the classical 
approximation of the system will consist of the portion of the 
black hole between the two ``$r={\rm const}$" surfaces. 
According to the analysis above, the entropy of this system vanishes. 
This is a rather intriguing result when one considers the fact that 
the boundary elements $B$ and $B'$ need not lie in the same wedge 
of the Kruskal diagram \cite{Erik}. If $B$ lies in, say, the right wedge 
and $B'$ lies in the left wedge, the system (in the classical 
approximation) will contain a black hole, yet the entropy 
will vanish. (Note that, in order for the action (9) and the path 
integral to be well defined, it is not necessary for the  spatial 
slices of the classical approximation to  be nonintersecting.) 
This, of course, is not the usual result for black hole entropy. In the 
usual path integral derivation of black hole entropy  
only the data on a single outer boundary element are specified 
\cite{QLEandMFI}. Loosely speaking, the information present in the 
inner boundary data has eliminated the black hole entropy. 

Consider, then, our system with inner and outer boundary elements, but 
with data specified only on the outer boundary element. Let us assume 
that the outer boundary data are chosen such that the system is classically 
approximated by a stationary, nonextreme black hole. The entropy of 
the system is the logarithm of the number of outer boundary states 
with metric $\sigma_{ab}$ and with $\varepsilon$ and 
$j_i$ in an interval $\Delta$. The number of such states 
is given by the  ``density of outer boundary states", 
\begin{equation}
  D[\varepsilon,j;\sigma] = \int d\varepsilon' \, dj' \, d\sigma' 
   \,\mu[\sigma'] 
  \, D[\varepsilon,j;\sigma|\varepsilon',j';\sigma'] \ ,\eqnum{10}
\end{equation} 
obtained by summing over all inner boundary states. Equation (8) shows 
that the integration over state labels $\varepsilon'$ and $j_i'$ can 
be viewed as a transformation to the canonical ensemble with 
$\beta'=0$ and $\beta' {v'}^i =0$. (One can show that ${v'}^i = 0$ 
as well \cite{EntropyInGeneral}.)  Thus, the boundary element $B'$ 
becomes an infinite temperature surface. For the black hole that 
approximates the system, $B'$ is  the
bifurcation surface of the Killing horizon. 

In Eq.~(10), which is the analog of Eq.~(3) with respect to the inner 
boundary data, a measure factor $\mu[\sigma']$ appears. This measure can be 
specified as follows. Consider the path integral for 
$D[\varepsilon,j;\sigma|\varepsilon',j';\sigma']$, constructed from the action 
$S^{ds}$. In the steepest descents approximation, the integrals in Eq.~(10) 
yield the condition that the variation $\delta (S^{ds} + \ln\mu[\sigma'])$ 
with respect to $\varepsilon'$, $j_i'$, and $\sigma_{ab}'$ should vanish. 
Equations~(6), (7), and (9) show that $\delta (S^{ds} + \ln\mu[\sigma'])$ 
includes the inner boundary terms 
\begin{eqnarray} 
    & &i\int dt \int_{B'} d^2x \left\{
     N\delta(\sqrt{\sigma}\varepsilon) - V^i\delta(\sqrt{\sigma}j_i) 
    \right.\nonumber\\ 
   & &+ \left. \sqrt{\sigma}( Ns^{ab}/2)\delta\sigma_{ab}\right\} +
   \int_{B'} d^2x \left( \delta\ln\mu/\delta\sigma_{ab} \right)
     \delta\sigma_{ab} \ .\eqnum{11} 
\end{eqnarray} 
From the first two terms we recover the results $\beta'=0$ and 
$\beta'{v^i}'=0$. From the remaining terms, and the definitions of 
$(Ns^{ab}/2)$ and $\beta$, we obtain the condition 
$\sqrt{\sigma} (-n^i\partial_i \beta){\sigma}^{ab} 
\approx 16\pi( \delta\ln\mu[\sigma]/\delta\sigma_{ab})$. Now, the factor 
$-n^i\partial_i \beta$, which is the inverse temperature gradient at $B'$, 
obeys $-n^i\partial_i \beta = 2\pi$. This is a universal result, in the  
sense that it applies to all bifurcate  Killing horizons in all theories of  
gravity. It can be derived, for example,  by considering  the response of
a  particle detector that maintains a constant proper distance from the 
horizon.\footnote{The boundary data are stationary and nonsingular, so I 
will assume that the
appropriate quantum state is the stationary Hadamard vacuum \cite{Wald}. 
Note that such a state exists only if the Killing vector field is 
everywhere timelike. For stationary, asymptotically flat black holes such 
as the Kerr solution, one  can possibly define the
Hadamard vacuum by taking the  outer  boundary $B$ to lie inside the 
speed--of--light surface. Also note that 
$-n^i\partial_i\beta$ equals $i\kappa P$, where $\kappa$ is the surface 
gravity and $P$ is the coordinate time period \cite{EntropyInGeneral}. If 
the black hole is asymptotically flat, and if the lapse function is 
normalized to unity at infinity, then $iP$ equals the inverse temperature
$\beta_\infty$ at  infinity. Under these conditions the result 
$-n^i\partial_i \beta = 2\pi$ is equivalent to $\beta_\infty = 2\pi/\kappa$.} 
Derived in this way, it is apparent  that the result does not  depend on the 
particular theory of gravity under consideration. Also it  does not depend 
on the  details of the spacetime geometry since,  at  sufficiently close 
distances, any bifurcate Killing horizon is physically  indistinguishable 
from a  ``Rindler horizon" (the horizon in flat spacetime  generated by a 
Lorentz boost). To be specific, then, the desired result  can be derived 
from the expression $\beta=2\pi/a$ for the inverse temperature  measured by 
a  detector undergoing constant acceleration $a$ in the Minkowski  vacuum of 
flat spacetime \cite{Wald}. It is not difficult to show that $1/a$ measures 
proper radial distance from the bifurcation surface. Hence, we have 
$-n^i\partial_i \beta = 2\pi$ and the condition on the measure $\mu[\sigma]$ 
becomes $\delta\ln\mu[\sigma]/\delta\sigma_{ab} 
\approx \sqrt{\sigma}{\sigma}^{ab}/8$.  
Integrating this expression, we find that to leading order 
$\ln\mu[\sigma'] \approx \int_{B'} d^2x \sqrt{\sigma}/4$. 

With  the measure $\mu[\sigma']$ and the previous result
$D[\varepsilon,j;\sigma|\varepsilon',j';\sigma'] \approx 1$,  the density 
of outer boundary states (10) becomes $D[\varepsilon,j;\sigma] \approx 
\exp(A/4)$. Here, $A = \int_{B'} d^2x \sqrt{\sigma}$ is the horizon area 
of the black hole that approximates the system. The entropy is 
then ${\cal S}[\beta,v;\sigma] \approx \ln D[\varepsilon,j;\sigma] \approx 
A/4$ as expected. 

The above reasoning can be used to derive the black hole entropy in any 
diffeomorphism invariant theory of  gravitational and matter  fields. Start 
with the general action ${S}$ of Refs.~\cite{IandW,EntropyInGeneral}. $S$ 
is the integral over spacetime ${\cal M}$ of a Lagrangian density 
${\cal L}$. Assume that a boundary term can  be added to $S$ to yield an 
action $S^{pf}$ whose fixed boundary data include  the induced metric on 
$\partial{\cal M}$. The path integral with action $S^{pf}$  is the 
partition function. As in Eq.~(7), the  variation of $S^{pf}$ defines 
the state labels $\varepsilon$, $j_i$, as well as the spatial stress 
$s^{ab}$. As emphasized in  Ref.~\cite{EntropyInGeneral}, the action
$S^{ds}$ for the density of states contains no boundary  terms when 
written in Hamiltonian form. Thus, $\delta S^{ds}$  contains no variations 
of $N$ or $V^i$ on $\partial\Sigma$, and $S^{ds}$ must differ from 
$S^{pf}$ by  boundary terms that change the fixed boundary data from 
$N$, $V^i$, and $\sigma_{ab}$ to $\varepsilon$, $j_i$, and $\sigma_{ab}$. 
In particular, $\delta S^{ds}$ must include boundary terms of the same 
form  as those that appear in the first integral of Eq.~(11). 

Black hole 
entropy is derived by summing over inner boundary states to form the 
density of outer boundary states, as in  Eq.~(10).  Integration over 
$\varepsilon'$ and $j'_i$ implies that the inner boundary  element $B'$ 
is the bifurcation surface of a black hole horizon. Integration over  
other state labels 
implies that the products of $\beta$ with certain ``intensive" variables  
should vanish at $B'$ \cite{EntropyInGeneral}. The measure 
$\mu[\sigma',\ldots]$ depends on the surface metric $\sigma'_{ab}$ and 
other parameters. It is obtained, as in Eq.~(11), from the requirement 
that $\delta (S^{ds} +\ln\mu)$ should vanish for variations in 
$\sigma_{ab}'$.  The spatial stress $(Ns^{ab}/2)$ is needed for this 
calculation, and is obtained from the following considerations. The 
actions $S$ and $S^{ds}$ differ by boundary terms. In 
Ref.~\cite{EntropyInGeneral} it is shown that one of these 
boundary terms has integrand 
$-4\sqrt{\sigma} n_i  U_0^{\perp ij\perp} \partial_j N$, and the other 
boundary terms are linear in the (undifferentiated) lapse $N$ or
shift $V^i$. Here, $U_0^{\mu\nu\rho\sigma}$ is the variational derivative 
of ${\cal L}$ with respect to the Riemann tensor 
${\cal R}_{\mu\nu\rho\sigma}$. Also, $\perp$ designates a component in the 
$u_\mu$ direction, where $u_\mu$ is orthogonal to $\Sigma$. Now,
$\delta S$  contains only boundary terms at $\partial{\cal M}$. If the 
lapse on an element of $\partial\Sigma$, say $B'$, vanishes, then $B'$ 
is a surface where the foliation degenerates. In this case the 
only possible contribution at $B'$ to the boundary terms of 
$\delta S$ is a  ``corner" 
term proportional to $\delta(\partial_i N)$ \cite{BoostedQLE}. On the 
other hand, we see that $\delta (S-  S^{ds})$ includes a term 
$\int_{B'}(-2\sqrt{\sigma}n_i  U_0^{\perp ij\perp} 
\partial_j N){\sigma}^{ab}\delta\sigma_{ab} $. 
This is not a corner term, so it must come from $\delta S^{ds}$. 
Therefore the spatial stress, which appears in the boundary terms 
of $\delta S^{ds}$, must be given by $(Ns^{ab}/2) 
= (2 n_i U_0^{\perp ij\perp}\partial_jN)\sigma^{ab} +\cdots$. Here, 
the unwritten terms contain undifferentiated linear factors of $N$ 
or $V$. The condition on the measure that follows from Eq.~(11) is 
then $2\sqrt{\sigma} n_i U_0^{\perp ij\perp} n_j (-n^k\partial_k\beta) 
{\sigma}^{ab} = \delta \ln\mu[\sigma,\ldots]/\delta\sigma_{ab}$, 
where $\partial_iN $ 
is proportional to $n_i$ \cite{EntropyInGeneral}. With the universal 
result $-n^k\partial_k \beta = 2\pi$ for the inverse temperature gradient, 
we find that to leading order the measure is given by 
$\ln\mu[\sigma',\ldots] \approx 8\pi \int_{B'} \sqrt{\sigma} 
n_iU_0^{\perp ij\perp} n_j$. 
This yields  the desired result \cite{IandW,EntropyInGeneral}, 
${\cal S}[\beta,v;\sigma]\approx\ln D[\varepsilon,j;\sigma] \approx  
-2\pi\int_{B'} \sqrt{\sigma} \epsilon_{\mu\nu}
U_0^{\mu\nu\rho\sigma}\epsilon_{\rho\sigma} $, where $\epsilon_{\mu\nu} 
= 2u_{{\scriptscriptstyle [}\mu} n_{\nu{\scriptscriptstyle ]}}$ is the
binormal of the  bifurcation surface $B'$. 

I would like to thank J.W.~York for valuable insights. 

 

\begin{references} 
\bibitem{Bstates} See A.P.~Balachandran, L.~Chandar, and A.~Momen, Nucl. 
Phys. B {\bf 461} 581 (1996); 
J.C.~Baez, J.P.~Muniain, and D.D.~P\'{\i}riz, Phys. Rev. D {\bf 52}, 
6840 (1995); 
L. Smolin, J. Math. Phys. {\bf 36}, 6417 (1995); 
C. Teitelboim, Phys. Rev. D {\bf 53}, 2870 (1996); 
S.~Carlip, Phys. Rev. D {\bf 55}, 878 (1997); 
M.~Ba\~{n}ados and A.~Gomberoff (gr-qc/9611044); 
and references therein. 
Results that are in conflict with this idea have been obtained by 
J.~Gegenberg, G. Kunstatter, and T. Strobl (gr-qc/9612033). 
\bibitem{QLEandMFI} J.D. Brown and J.W. York, Phys. Rev. D {\bf 47}, 
1407 (1993); {\bf 47}, 1420 (1993). 
\bibitem{Dbranes} 
See A.~Strominger and C.~Vafa, Phys. Lett. {\bf 379B}, 99 (1996); 
G.~Horowitz (hep-th/9604051); and references therein.  
\bibitem{IandW} V.~Iyer and R.M.~Wald, Phys. Rev. D {\bf 50}, 846 (1994). 
\bibitem{EntropyInGeneral} J.D.~Brown, Phys. Rev. D {\bf 52}, 7011 (1995). 
\bibitem{SachHill} T.L.~Hill, {\em Statistical Mechanics} (McGraw-Hill, 
New York, 1956); R.A.~Sach, Mol. Phys. {\bf 2}, 8 (1959).  
\bibitem{BoostedQLE} 
J.D. Brown, S.R. Lau, and J.W. York,   in preparation. 
\bibitem{BMY} 
J.D. Brown, E.A. Martinez, and J.W. York, Phys. Rev. Lett. {\bf 66}, 
2281 (1991). 
\bibitem{Erik} 
E.A. Martinez, Phys. Rev. D {\bf 51}, 5732 (1995). 
\bibitem{Wald} 
R.M.~Wald, {\em Quantum Field Theory in Curved Spacetime and Black 
Hole Thermodynamics} (University of Chicago Press, Chicago, 1994). 
\end{references}
\end{document}